\documentclass{appolb} 
\usepackage{graphicx}
\usepackage[utf8]{inputenc}
\usepackage{bm}
\usepackage{amsmath}
\global\long\def\aB{a_{{\scriptscriptstyle \text{B}}}}%
\global\long\def\vk{\bm{k}}%
\global\long\def\vq{\bm{q}}%
\global\long\def\vr{\bm{r}}%
\global\long\def\vnabla{\bm{\nabla}}%
\global\long\def\d{\mbox{d}}%

\begin{document}
\title{Energy-momentum tensor of a hydrogen atom: stability, $D$-term,
and the Lamb shift%
 \thanks{Talk presented at {\em XXIX Cracow Epiphany Conference on Physics at the Electron-Ion Collider and Future Facilities, Cracow, Poland, 16-19 January 2023}}}

\author{Andrzej Czarnecki
\address{Department of Physics, University of Alberta, Edmonton,
  Alberta T6G 2E1, Canada}
\\[3mm]
{Yizhuang Liu
\address{Institute of Theoretical Physics, Jagiellonian University,
  30-348 Krak\'ow, Poland}
}
\\[3mm]
{Syed Navid Reza
\address{Department of Physics, University of Alberta, Edmonton,
  Alberta T6G 2E1, Canada}
}}
\preprint{Alberta Thy 7-23}
\maketitle
\begin{abstract}
We clarify two issues related to the so-called $D$-term in matrix
elements of the energy-momentum tensor. First, we show that in a
stable system, the $D$-term can have either sign, contrary to claims
that it must be negative. Second, we demonstrate a logarithmic
enhancement of the $\mathcal{\alpha}$ correction to the $D$-term in
any state of the hydrogen atom. We contrast this enhancement with the
Lamb shift where it is present only in S-states. 
\end{abstract}

\section{Introduction}

Gravitational interaction of a particle is governed by its energy-momen\-tum
tensor (EMT), $T^{\mu\nu}$. Although gravitational interactions are
too weak for present experiments to probe them, distributions of mechanical
properties encoded in the EMT, such as the energy density, shear,
and pressure, can be revealed in scattering processes. For example,
a very recent study \cite{Duran:2023} determined the gluonic contribution
to gravitational form factors (GFF) of the proton.

The GFFs were introduced in~\cite{Kobzarev:1962wt}, and were
subsequently studied in \cite{Pagels:1966zza}. Radiative corrections
to graviton-matter interaction were studied in \cite{Milton:1976jr}.
Recently, GFF have attracted attention because indirect measurements
are possible through generalized parton distributions (GPD) \cite{Ji:2004gf}
in processes like deeply virtual Compton scattering (DVCS) \cite{Ji:1996nm}.
Experiments like the measurements of the gluonic and quark contributions to the GFFs \cite{Duran:2023,Burkert:2018bqq} are ongoing in JLab and are planned in the future Electron-Ion Collider in Brookhaven.

For a spin-0 particle without internal structure the matrix element
of $T^{\mu\nu}$ between states with momenta $p_{1}=p+\frac{q}{2}$,
$p_{2}=p-\frac{q}{2}$, is (we use $\hbar=c=1$) 
\begin{equation}
\left\langle p_{2}\left|T^{\mu\nu}\left(x\right)\right|p_{1}\right\rangle =\left[2p^{\mu}p^{\nu}-\frac{1}{2}\left(q^{\mu}q^{\nu}-q^{2}g^{\mu\nu}\right)\right]e^{i\left(p_{2}-p_{1}\right)x}.\label{eq:1}
\end{equation}
If the internal degrees of freedom of the system become relevant at
accessible energies, the two tensor structures are modified by $q^{2}$-dependent
form factors (see for example Ref. \cite{Polyakov:2018zvc}),
\begin{equation}
\left\langle p_{2}\left|T^{\mu\nu}\left(x\right)\right|p_{1}\right\rangle =\left[A\left(q^{2}\right)p^{\mu}p^{\nu}+\frac{1}{2}D\left(q^{2}\right)\left(q^{\mu}q^{\nu}-q^{2}g^{\mu\nu}\right)\right]e^{i\left(p_{2}-p_{1}\right)x}.\label{eq:2}
\end{equation}
In the present paper, we are particularly interested in the $D$-term
\cite{Polyakov:2018zvc}, which we discuss with the example of a spin-0
system. Examples include a pion (discussed in this conference \cite{Broniowski:2023his})
or a hydrogen atom. 

Comparing Eqs.~\eqref{eq:1} and \eqref{eq:2}, we see that $D\equiv D\left(q^{2}=0\right)=-1$.
It has been conjectured that in any stable system, the $D$ must be
negative \cite{Perevalova:2016dln}. This is indeed the case for Q-balls
and Q-clouds studied in \cite{Mai:2012yc,Cantara:2015sna}, for a
liquid drop model considered in \cite{Polyakov:2018zvc}, for the
bag model of the nucleon in \cite{Ji:1997gm}, and for the chiral
quark soliton model \cite{Wakamatsu:2007uc}. The $D$-term for spin-0
point-like and composite particles is discussed in \cite{Hudson:2017oul}
and the dynamic origin of the $D$-term for a spin-1/2 fermion in
\cite{Hudson:2017oul}. The $D$-term for pion in relativistic theory
is calculated in \cite{Krutov:2020ewr}.

However, Ref.~\cite{Ji:2021mfb} provided a counterargument that
the $D$-term can also be positive without endangering mechanical
stability. In Section \ref{sec:von-Laue's-stability} we review Max
von Laue's stability criterion and discuss its connection with the
sign of the $D$-term. In Section \ref{sec:The-energy-momentum-tensor}
we discuss this sign with an example taken from classical mechanics.

In Sections \ref{sec:Lamb} and \ref{sec:LogD} we focus on the hydrogen
atom since it can be studied analytically. In particular, it has been
found that radiative effects in $D\left(q^{2}\right)$ are enhanced
logarithmically \cite{Ji:2021mfb,Ji:2022exr}. In Section \ref{sec:Lamb}
we review the logarithmic Lamb shift correction. Finally, in Section
\ref{sec:LogD} we contrast it with the logarithmic correction to
the $D$-term.

\section{von Laue's stability criterion\label{sec:von-Laue's-stability}}

In a static system, the conservation law $\partial^{\mu}T_{\mu\nu}=0$
has only spatial derivatives, 
\begin{equation}
\nabla^{i}T_{i\nu}=0.\label{stab1}
\end{equation}
Momentum density $T_{i0}$ vanishes in a static system so we consider
only $\nu=j$. Max von Laue proposed \cite{Laue:1911lrk,MvL1} to
integrate Eq.~\eqref{stab1} over a surface consisting of a cross-section $\sigma$ of the static system and closing far away, where
$T_{ij}$ is assumed to vanish. We repeat here von Laue's reasoning
that lead him to formulate a criterion of stability of a system.

We work in the system's rest frame. One finds that the following integral
over the cross-section vanishes, 
\begin{equation}
\int_{\sigma}T^{ij}n_{j}\d\sigma=0,\label{int_sigma}
\end{equation}
where $n_{j}$ are the components of a vector normal to the cross-section. The closing part of the surface does not contribute to the
integral. Eq.~\eqref{int_sigma} becomes a system of three equations
for $i=x,y,z$, 
\begin{align}
\int T^{xi}\d y\d z=0.
\end{align}
We integrate this over all $x$, 
\begin{equation}
\int T^{xi}\d^{3}r=0.\label{stability}
\end{equation}
We conclude that each component $T^{xi}$, and in general each $T^{ij}$,
$i,j=x,y,z$, integrated over the whole volume of the system gives
zero. This is von Laue's stability condition.

As a non-trivial application, consider a drop of water in vacuum in
conditions of weightlessness \cite{Polyakov:2018zvc}. Due to surface
tension $\sigma$, there is pressure inside the drop, $p(r)=2\sigma/R$
for $r<R$, where $R$ is the radius of the drop (this is Laplace's
formula \cite{LL6}). Pressure $p$ is a diagonal element $T^{ii}$
(in an isotropic system all three such elements are the same). The
stability condition \eqref{stability} tells us that 
\begin{equation}
\int p(r)\d^{3}r=4\pi\int_{0}^{R}p(r)r^{2}\d r=0.\label{drop}
\end{equation}
This vanishing of the integral is of course only possible if $p(r)$
is negative somewhere. There is a thin surface layer where surface
tension makes $T^{ii}$ negative precisely to such an extent that it
cancels the positive contribution of the bulk. This tension holds
the drop together.

On the other hand, the $D$-term can be expressed as \cite{Polyakov:2018zvc}
(we assume spherical symmetry)
\begin{equation}
D=4\pi m\int p\left(r\right)r^{4}\d r,
\end{equation}
which has two additional powers of $r$ in comparison with Eq.~\eqref{drop}.
This gives larger weight to $p\left(r\right)$ at larger values of
$r$. If, as in the case of a liquid drop, the negative contribution
comes from the outer boundary, a negative $D$ results. On the other
hand, other binding mechanisms may have negative $p$ at short distances,
leading to a positive $D$. This is illustrated in the next Section.

\section{The energy-momentum tensor for a classical system \label{sec:The-energy-momentum-tensor}}

A negative $D$-term is unusual in classical mechanics. To demonstrate
this we consider the system composed of a static charge $e=\sqrt{4\pi\alpha}>0$
at $r=0$ (a nucleus) and a negatively charged point-like particle
in circular motion around it with radius $R$ and velocity $mv^{2}=\frac{\alpha}{R}$
(we use such units that $\hbar=c=\epsilon_{0}=1$). The EMT of the
system reads 
\begin{align}
T^{ij}(\vec{r})=mv^{i}v^{j}\delta^{3}\left(\vec{r}-\vec{x}(t)\right)-E^{i}E^{j}+\frac{\delta^{ij}}{2}\vec{E}^{2}\ ,
\end{align}
where the electric field is approximated by the static Coulomb fields
of the nucleus and of the orbiting particle, 
\begin{align}
\vec{E}\approx\frac{e}{4\pi}\frac{\vec{r}}{r^{3}}-\frac{e}{4\pi}\frac{\vec{r}-\vec{x}(t)}{|\vec{r}-\vec{x}(t)|^{3}}\equiv\vec{E}_{p}+\vec{E}_{e}\ .
\end{align}
We first show that $\int d^{3}\vec{r}\ T^{ii}(\vec{r})=2T+V=0$. Indeed,
\begin{align}
\int d^{3}\vec{r}\ T^{ii}(\vec{r})=mv^{2}+\frac{1}{2}\int d^{3}\vec{r}\vec{E}^{2}\ ,
\end{align}
where 
\begin{align}
\frac{1}{2}\int d^{3}\vec{r}\ \vec{E}^{2}=\frac{1}{2}\int d^{3}\vec{r}\left(\vec{E}_{e}^{2}+\vec{E}_{p}^{2}\right)+\int d^{3}\vec{r}\ \vec{E}_{e}\cdot\vec{E}_{p}\ .
\end{align}
The self-energy contributions vanish in dimensional regularization
(DR), while the last integral simply gives the Coulomb potential,
\begin{align}
\int d^{3}\vec{r}\ \vec{E}_{e}\cdot\vec{E}_{p}=-\frac{\alpha}{|\vec{x}(t)|}=-\frac{\alpha}{R}.
\end{align}
Therefore one simply has 
\begin{align}
\int d^{3}\vec{r}\ T^{ii}(\vec{r})=mv^{2}-\frac{\alpha}{R}=2T+V=0\ ,
\end{align}
equivalent to the virial theorem. We now move to the $\int d^{3}\vec{r}\ r^{2}T^{ii}$
integral. We first consider 
\begin{align}
I=\int d^{3}\vec{r}r^{2}\bigg(\frac{1}{r^{4}}+\frac{1}{|\vec{r}-\vec{R}|^{4}}-\frac{2r^{2}-2\vec{r}\cdot\vec{R}}{r^{3}|\vec{r}-\vec{R}|^{3}}\bigg)\ ,
\end{align}
using $r^{2}=R^{2}+|\vec{r}-\vec{R}|^{2}+2\vec{R}\cdot(\vec{r}-\vec{R})$
in the second term and $r^{2}-\vec{r}\cdot\vec{R}=|\vec{r}-\vec{R}|^{2}+\vec{R}\cdot(\vec{r}-\vec{R})$
in the third term, one has 
\begin{align}
I & =\int d^{3}\vec{r}\bigg(\frac{1}{r^{2}}+\frac{1}{|\vec{r}-\vec{R}|^{2}}+\frac{R^{2}}{|\vec{r}-\vec{R}|^{4}}+\frac{2\vec{R}\cdot(\vec{r}-\vec{R})}{|\vec{r}-\vec{R}|^{4}}\nonumber \\
 & \qquad-\frac{2}{|\vec{r}||\vec{r}-\vec{R}|}+\frac{2\vec{R}\cdot(\vec{r}-\vec{R})}{r|\vec{r}-\vec{R}|^{3}}\bigg)\ .
\end{align}
The first four terms all vanish in DR, leaving only 
\begin{align}
I=I_{1}+I_{2}\equiv\int d^{3}\vec{r}\bigg(-\frac{2}{|\vec{r}||\vec{r}-\vec{R}|}+\frac{2\vec{R}\cdot(\vec{r}-\vec{R})}{r|\vec{r}-\vec{R}|^{3}}\bigg)\ .
\end{align}
In DR the above can be further calculated as 
\begin{align}
I_{1}(D) & =-\frac{2}{\pi}\int\frac{d\alpha_{1}d\alpha_{2}}{\sqrt{\alpha_{1}\alpha_{2}}}\int d^{D}\vec{r}e^{-(\alpha_{1}+\alpha_{2})r^{2}-\frac{\alpha_{1}\alpha_{2}}{\alpha_{1}+\alpha_{2}}R^{2}}\nonumber \\
 & \rightarrow4\pi R|_{D=3}\ ,\\
I_{2}(D) & =-2R^{2}\frac{2}{\sqrt{\pi}}\frac{1}{\sqrt{\pi}}\int\frac{d\alpha_{1}d\alpha_{2}\sqrt{\alpha_{1}\alpha_{2}}}{\alpha_{1}+\alpha_{2}}\int d^{D}\vec{r}e^{-(\alpha_{1}+\alpha_{2})r^{2}-\frac{\alpha_{1}\alpha_{2}}{\alpha_{1}+\alpha_{2}}R^{2}}\nonumber \\
 & \rightarrow-4\pi R\ .
\end{align}
Therefore, one simply has 
\begin{align}
\int d^{3}\vec{r}\ r^{2}\,T^{ii}(\vec{r})=mv^{2}R^{2}+\frac{e^{2}}{32\pi^{2}}\bigg(I_{1}+I_{2}\bigg)=\alpha R\ ,
\end{align}
or $\tau=\frac{1}{12}\int d^{3}\vec{r}\ r^{2}\,T^{ii}(\vec{r})=\frac{\alpha R}{12}>0$.
If one uses $R\sim\langle r\rangle=\frac{3}{2\alpha m}$, then $\tau\sim\frac{1}{8m}$,
comparing to the leading order (LO) result $\tau_{H}=\frac{1}{4m}$ (see Eq.~(61) in Ref.~\cite{Ji:2022exr}).

\section{Logarithmic enhancement of $\mathcal{\alpha}$ corrections to the
Lamb shift\label{sec:Lamb}}

In this Section we summarize Welton's heuristic explanation of the
leading part of the Lamb shift \cite{Welton:1948zz}.

Welton argued that, although the expectation value of the electromagnetic
field strength vanishes in vacuum, there are non-zero fluctuations
such that $E_{k}^{2}=\left\langle 0\left|E_{\vk}^{2}\right|0\right\rangle \neq0$
for all modes, where $\vk$ denotes a wave vector. Consider the system
to be enclosed in a large cube of volume $V$, with periodic boundary
conditions. Since the vacuum energy of one mode is $\omega_{k}/2$,
where $\omega_{k}=\left|\vk\right|$, and on the other hand the energy
density is related to the squares of the electric and the magnetic
fields, we obtain 
\begin{equation}
E_{k}^{2}=\frac{\omega_{k}}{2V}.
\end{equation}
$E_{\vk}$ is the amplitude of a plane wave, $E_{\vk}\exp i\vk\cdot\vr$.
The mode density of such running plane waves is 
\begin{equation}
\frac{V\d^{3}k}{\left(2\pi\right)^{3}}\label{LambWelton:density}
\end{equation}

Because of the fluctuating electric field, the electron's position
is modified and this displacement in the Coulomb potential $U_{c}\left(r\right)=-\alpha/r$
gives rise to an extra effective potential $\delta U$, 
\begin{align}
\left\langle U_{c}\left(\vr+\vq\right)\right\rangle  & =U_{c}\left(r\right)+\underbrace{\left\langle \vq\right\rangle }_{0}\cdot\vnabla U_{c}+\frac{1}{2}\underbrace{\left\langle q^{i}q^{j}\right\rangle }_{\frac{\delta_{ij}}{3}\left\langle q^{2}\right\rangle }\nabla^{i}\nabla^{j}U_{c}+\dots,\\
\delta U & =\left\langle U_{c}\left(\vr+\vq\right)\right\rangle -U_{c}\left(r\right)\simeq\frac{1}{6}\left\langle q^{2}\right\rangle \nabla^{2}U_{c}=\frac{1}{6}\left\langle q^{2}\right\rangle \alpha4\pi\delta^{3}\left(\vr\right).\label{LambWelton:dU}
\end{align}
In order to find the mean disturbance $\left\langle q^{2}\right\rangle $,
Welton considered the equation of motion, assuming a free electron,
\begin{equation}
m\ddot{q}=-eE
\end{equation}
where $E$ is the electric field. In the Fourier space, for each mode
$\vk$, 
\begin{equation}
-m\omega^{2}q_{k}=-eE_{k}\to q_{k}=\frac{e}{m\omega^{2}}E_{k}.
\end{equation}
The total disturbance, summed over all modes, including a factor 2
for two polarizations, 
\begin{align}
\left\langle q^{2}\right\rangle  & =2\int\left(\frac{e}{m\omega^{2}}\right)^{2}\frac{V\d^{3}k}{\left(2\pi\right)^{3}}E_{\vk}^{2}\\
 & =\frac{2\alpha}{\pi m^{2}}\int\frac{\d k}{k}.
\end{align}
For small $k$, this integral is cut off by energies on the order
of excitations of the atom, where the electron cannot be treated as
free: $k_{\min}\sim\alpha^{2}m$. For large $k$, it is cut off at
the inverse Compton wavelength of the electron: when the electron
absorbs a large momentum it becomes relativistic and its increased
inertia decreases its displacement. The upper limit, $k_{\max}\sim m$,
does not contain $\alpha$. So the logarithmic dependence on $\alpha$
is determined by the lower limit, 
\begin{equation}
\left\langle q^{2}\right\rangle =\frac{2\alpha}{\pi m^{2}}\ln\frac{1}{\alpha^{2}}+\text{non-logarithmic terms.}\label{LambWelton:q2}
\end{equation}
Substituting this into \eqref{LambWelton:dU}, we find 
\begin{equation}
\delta U=\frac{8}{3}\frac{\alpha^{2}}{m^{2}}\ln\frac{1}{\alpha}\cdot\delta^{3}\left(\vr\right)
\end{equation}
For example, in the 2S state, $\left\langle \delta^{3}\left(\vr\right)\right\rangle _{2S}=\psi_{2S}^{2}\left(0\right)=1/8\pi\aB^{3}$
where $\aB$ is the Bohr radius. That state's energy changes by 
\begin{align}
\left\langle \delta U\right\rangle _{2S} & =\frac{m}{3\pi}\alpha^{5}\ln\frac{1}{\alpha}.
\end{align}
This energy corresponds to the frequency of about 1000 MHz, which
is the shift observed by Lamb and Retherford \cite{Lamb:1947zz}.
We stress that the perturbing potential $\delta U$ in Eq.~\ref{LambWelton:dU}
is proportional to the Laplacian of the Coulomb potential, which vanishes
except where the electric charge is present, that is in the nucleus.
For this reason this mechanism, giving rise to the logarithmic enhancement,
applies only to S-states (vanishing angular momentum).

\section{Logarithmic correction to $D$-term and effective theory\label{sec:LogD}}

Contrary to the logarithmic correction to the Lamb shift, which
depends on $\nabla^{2}V$, the NLO logarithm for the $D$-term is
almost universal. Indeed, the logarithmically-enhanced contribution
reads
\begin{align}
D_{{\rm NLO}}=\frac{\alpha}{6\pi}\sum_{M}\frac{2\vec{v}_{0M}\cdot\vec{v}_{M0}}{D(E_{M}-E_{0})}\bigg(\ln\frac{4(E_{M}-E_{0})^{2}}{m_{e}^{2}}-\frac{1}{4}\bigg)\ ,
\end{align}
with the coefficient of the logarithm,
\begin{align}
\sum_{M}\frac{2\vec{v}_{0M}\cdot\vec{v}_{M0}}{D(E_{M}-E_{0})}\equiv\frac{1}{m_{e}}\ ,\label{eq:sumrule}
\end{align}
being independent of any details of the bound state. To some extent,
Eq.~(\ref{eq:sumrule}) simply reflects the fact that in $D$-dimension,
the mass dimension of any normalized wave function equals $\frac{D}{2}$.
Indeed, if one introduces the ``dilatation operator''
\begin{align}
\hat{D}=-i\vec{x}\cdot\vec{p}=-x^{\mu}\frac{\partial}{\partial x^{\mu}}\ ,
\end{align}
then it is easy to see that Eq.~(\ref{eq:sumrule}) is equivalent
to
\begin{align}
\langle0|\hat{D}|0\rangle=\frac{D}{2}\ ,
\end{align}
nothing but the mass dimension of the wave function $\langle x|0\rangle$.
This connection between the EMT form factor to the re-scaling property
of the wave function is expected, since to some extent $T^{ii}$ also
measures the response of the wave function under a spatial re-scaling
$\vec{x}\rightarrow\lambda\vec{x}$.

Another important fact that should be mentioned is that this logarithm,
although obtained through NRQED, matches precisely to the IR logarithms
in relativistic QED for a free-electron
\begin{align}
D_{{\rm QED}}\sim\frac{\alpha}{6\pi m_{e}}\bigg(\ln\frac{4Q^{2}}{m_{e}^{2}}-\frac{11}{12}\bigg)\ ,
\end{align}
providing a nice demonstration of the principle of the effective field
theory. To some extent, in the presence of ``criticality'', in the
sense that scale separations $\frac{\alpha m_{e}}{\alpha^{2}m_{e}},$
$\frac{m_{e}}{\alpha m_{e}}$ become large, non-trivial structures
with clean boundaries exists only near a small number of sharp peaks
in the logarithmic scale. In the intermediate energy scales such as
$\alpha m_{e}\ll\mu\ll m_{e}$, ``colored noises'', or self-similar
random fluctuations without clear shape/boundary, characterized by
simple scaling law possibly with logarithmic corrections, dominate.
This vast sea of noise, although not splendid at first glance, actually
serves as an amorphous ``bridge'' joining smoothly the otherwise
divided worlds in the IR and UV, witnessing the ``matching'' between
effective theories. And if you look deeper into the cloud, you see
beauties, such as the spin-spin correlator $G(z)$ for the two-dimensional Ising model as a function of separation $z$~\cite{Onsager:1943jn,Wu:1975mw,DiFrancesco:1997nk},
\begin{align}
G(z)\rightarrow\frac{1}{z^{\frac{1}{4}}}\ ,
\end{align}
or  the running coupling constant $g(L)$
for the four-dimensional critical Ising model at scale $L$~\cite{Callan:1970yg,Feldman:1987zq,Aizenman:2019yuo},
\begin{align}
g(L)\rightarrow\frac{16\pi^{2}}{3\ln L}  \ ,
\end{align}
lasting forever.

\section*{Acknowledgements}

This research was supported by the Natural Sciences and Engineering
Research Council of Canada (NSERC). 


\end{document}